\documentclass[sigconf,nonacm]{acmart}
\usepackage{enumitem}
\usepackage{multirow}

\AtBeginDocument{%
  \providecommand\BibTeX{{%
    \normalfont B\kern-0.5em{\scshape i\kern-0.25em b}\kern-0.8em\TeX}}}

\copyrightyear{2022} 
\acmYear{2022} 
\setcopyright{acmcopyright}\acmConference[SIGIR '22]{Proceedings of the 45th International ACM SIGIR Conference on Research and Development in Information Retrieval}{July 11--15, 2022}{Madrid, Spain}
\acmBooktitle{Proceedings of the 45th International ACM SIGIR Conference on Research and Development in Information Retrieval (SIGIR '22), July 11--15, 2022, Madrid, Spain}
\acmPrice{15.00}
\acmDOI{10.1145/xx}
\acmISBN{xx}

\begin{document}
\fancyhead{}

\title{Adversarial Filtering Modeling on Long-term User Behavior Sequences for Click-Through Rate Prediction}

\author{Xiaochen Li}
\thanks{X. Li and R. Zhong share the co-first authorship.}
\affiliation{%
  \institution{Alibaba Group}
  \city{Beijing}
  \country{China}
}
\email{xingke.lxc@alibaba-inc.com}

\author{Rui Zhong}
\affiliation{%
  \institution{Alibaba Group}
  \city{Beijing}
  \country{China}
}
\email{ruishi.zr@alibaba-inc.com}

\author{Jian Liang}
\affiliation{%
  \institution{Alibaba Group}
  \city{Beijing}
  \country{China}
}
\email{xuelang.lj@alibaba-inc.com}

\author{Xialong Liu}
\affiliation{%
  \institution{Alibaba Group}
  \city{Beijing}
  \country{China}
}
\email{xialong.lxl@alibaba-inc.com}

\author{Yu Zhang}
\affiliation{%
  \institution{Lazada Group}
  \city{Beijing}
  \country{China}
}
\email{daoji@lazada.com}

\begin{abstract}
Rich user behavior information is of great importance for capturing and understanding user interest in click-through rate (CTR) prediction. To improve the richness, collecting long-term behaviors becomes a typical approach in academy and industry but at the cost of increasing online storage and latency. Recently, researchers have proposed several approaches to shorten long-term behavior sequence and then model user interests. These approaches reduce online cost efficiently but do not well handle the noisy information in long-term user behavior, which may deteriorate the performance of CTR prediction significantly. To obtain better cost/performance trade-off, we propose a novel Adversarial Filtering Model (ADFM) to model long-term user behavior. ADFM uses a hierarchical aggregation representation to compress raw behavior sequence and then learns to remove useless behavior information with an adversarial filtering mechanism. The selected user behaviors are fed into interest extraction module for CTR prediction. Experimental results on public datasets and industrial dataset demonstrate that our method achieves significant improvements over state-of-the-art models.
\end{abstract}

\ccsdesc[500]{Information systems~Online advertising}

\keywords{Click-Through Rate Prediction; User Behavior Modeling; Long-Term User Behavior}

\maketitle

\section{Introduction}
Click through rate prediction plays a vital role in recommender system and online advertising. Due to the rapid growth of user historical behavior data, user behavior modeling has been widely adopted in CTR prediction, which focuses on capturing the dynamics of user interest from user historical behaviors~\cite{DIN,DIEN,2020ComicRec,bert4rec,DeepForCtr}. These models are mainly designed for short-term behavior sequence with limited length (e.g., less than 100). A natural extension is expanding the time window of historical behavior and incorporating more information. The length can be tens of thousands whereas the cost of storage and latency becomes rather high, especially for large-scale recommender system. How to design an efficient model for long-term user behavior has become a great challenge.

A straightforward solution is transforming long sequence to short sequence and applying classical user behavior modeling approaches directly. There has been some recent works trying to shorten long sequence~\cite{HPMN,MIMN,SIM}. MIMN~\cite{MIMN} proposes a memory network-based model for long sequential user behavior modeling. It maintains a fixed-length memory and incrementally updates the memory when new behavior arrives. A similar model is HPMN~\cite{HPMN} which uses memory network in lifelong sequential modeling. Memory network-based models have great advantages in saving storage and latency cost but fail to precisely capture user interest given a specific target item since encoding all historical behaviors into a fixed-size memory introduces massive noise. To overcome the limitations, SIM~\cite{SIM} designs a search-based model to selects relevant behaviors by hard search or soft search. Hard search selects sequence behaviors belonging to the category of target item. Soft search uses maximum inner product search to retrieval sequence behaviors similar to target item based on embedding vectors of items. SIM performs better than MIMN but at the risk of increasing online storage. Moreover, the selection strategies of SIM may not accurately identify relevant behaviors. For hard search, it is a rule-based strategy which causes information loss and may not remove noisy information efficiently. For soft search, it relies on embedding vectors of items to calculate the similarity between target item and sequence behaviors. However, embedding vectors are mainly learned for CTR prediction task and the similarity does not necessarily imply the relevance of sequence behaviors. The experimental results also show that soft search only performs slightly better than hard search~\cite{SIM}. UBR~\cite{UBR} organizes user history behaviors into a inverted index and generates search queries to retrieve relevant behaviors. It ranks the candidate behaviors using BM25 and feeds selected behaviors to an attention-based CTR model. Although interesting, the ranking function in UBR is independent with CTR model, which can not be end-to-end optimized and may yield sub-optimal performance in retrieving relevant behaviors. In summary, these models reduce online cost but does not well handle the noisy information in long-term user behavior, which may deteriorate the performance significantly.

To better understand the problem, we carefully examine cases of long-term user behavior sequence and find that there are two types of noise, including duplicate behaviors and useless behaviors. Duplicate behaviors can be popular items, brands, shops that user visits multiple times. Keeping duplicate behaviors does not bring much new information but makes the limited-length sequence dominated by several hot items. Useless behaviors can be items clicked accidentally or long-tail items. These behaviors may not reflect user recent interests and have little value in online prediction.

We propose a novel Adversarial Filtering Model to remove duplicate and useless behavior from long-term user behavior sequences with low cost of storage and latency. In specific, we first use a hierarchical aggregation representation to group duplicate behaviors, score and select the top-k useful behaviors, and then use multi-head attention to extract user interest from selected sequences. We also propose an adversarial filtering mechanism to encourage the selection of useful behaviors. 

The rest of the paper is organized as follows: Section 2 introduces our ADFM model. Section 3 presents experimental results on public datasets and industrial dataset. Section 4 concludes the paper.

\section{The proposed approach}
In this Section, we first introduce a base CTR model as benchmark. Then we detail model structure and optimization strategy of ADFM.

\subsection{Base CTR model}
{\bfseries Input features.} The features used in base CTR model include: (1) Item profile: Item id and its side information (e.g., brand id, shop id, category id). (2) User profile: User id, age, gender and income level. (3) Short-term user behavior: \sloppy For each user $u \in U$, there are several historical behavior sequences $ s_{t,v}^{u}=[b_{t,v}^{1},...,b_{t,v}^{i},...,b_{t,v}^{n}] $, where $b_{t,v}^{i}$ is the i-th behavior with behavior type $t \in \{impression, click, add\ to\ cart, pay\}$ and behavior target $v \in \{item, brand, shop, category\}$. The value of each behavior $b_{t,v}^{i}$ is the id of its behavior target, e.g., item id. The sequence is sorted by occurrence time of $b_{t,v}^{i}$. The time window is 3 days and the length does not exceed 100. (4) Long-term user behavior: The representation is the same as that of short-term one. The time window can be several months and the length can be tens of thousands.

{\bfseries Embedding layer.} Embedding layer is a common operation to transform high-dimensional sparse features to low-dimensional dense embedding. For sparse feature $f_{i}$, the corresponding embedding dictionary is denoted as $E_{i}=[e_{i}^{1},...,e_{i}^{j},...e_{i}^{N_{i}}] \in R^{D \times N_{i}}$, where $e_{i}^{j}\in R^{D}$ is an embedding vector, $D$ is the embedding dimension, ${N_{i}}$ is the number of sparse features. Embedding layer looks up the embedding table and produces a list of embedding vectors. If $f_{i}$ is a one-hot vector with the j-th element $f_{i}[j]=1$, the embedding vector list is $\{e_{i}^{j}\}$. If $f_{i}$ is a multi-hot vector with $f_{i}[j]=1$ for $ j \in \{i_{1}, i_{2}, ..., i_{k}\}$, the embedding vector list is $\{e_{i}^{i_{1}},e_{i}^{i_{2}},...,e_{i}^{i_{k}}\}$.

{\bfseries Pooling layer.} The embedding vector list is fed into pooling layer to get a fixed-length embedding vector. Sum pooling is adopted in our work. 

{\bfseries MLP layer.} All the embedding vectors of input features are concatenated and fed into MLP layer. MLP layer is used to learn the nonlinear interaction between features. The output of MLP layer is the probability of target item being clicked.

{\bfseries Loss function.} Negative log-likelihood function is used:
\begin{equation}
\label{eqn:ctr_loss}
    L = -\frac{1}{N}\sum_{(x,y)\in S}(ylogp(x)+(1-y)log(1-p(x)))
\end{equation}
where $S$ is the training set of size $N$, $x$ is input features, $y \in \{0,1\}$ is click label and $p(x)$ is the predicted CTR. 

\subsection{The ADFM model}
\begin{figure}[h]
  \centering
  \includegraphics[width=0.7\linewidth]{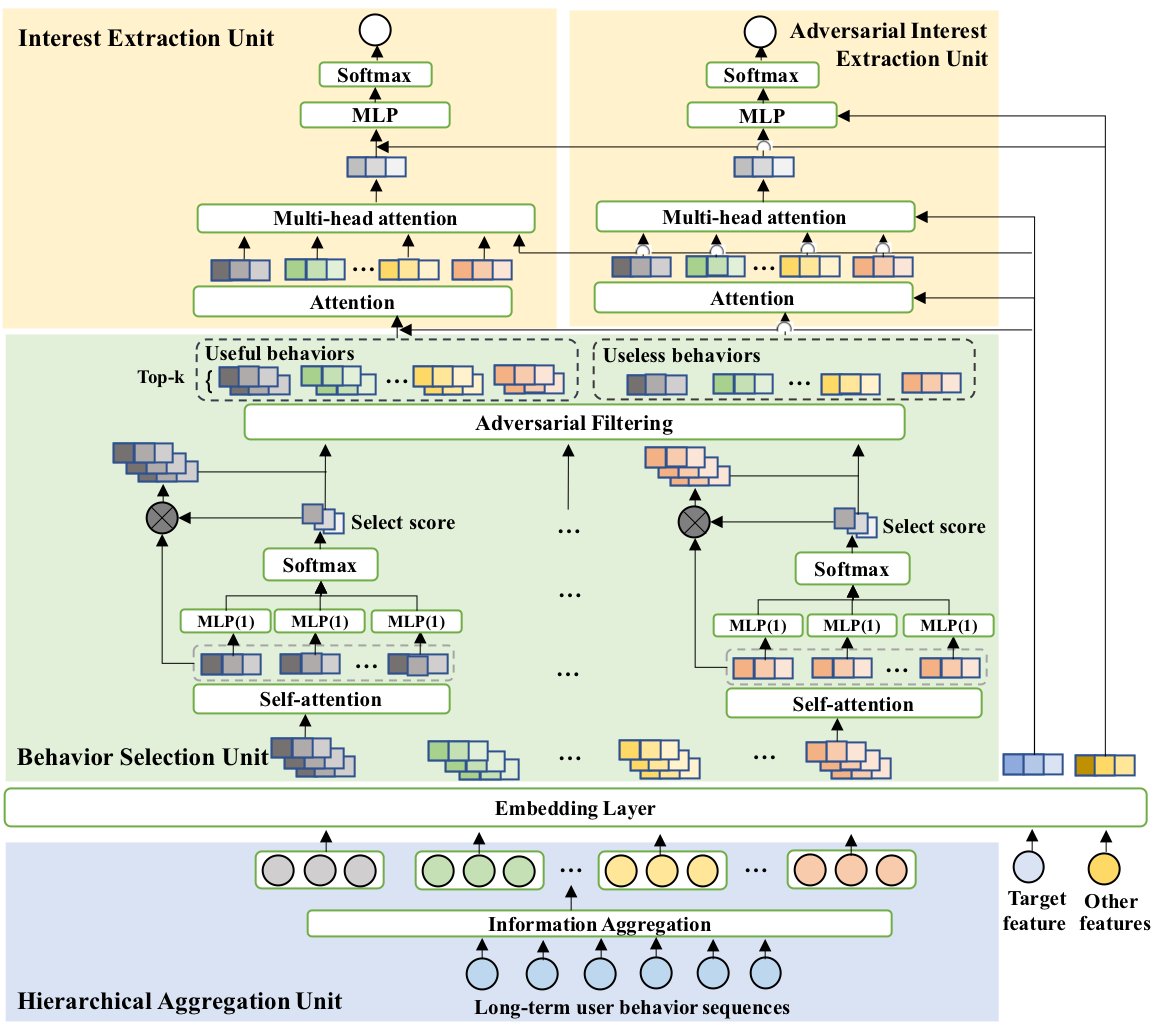}
  \caption{The structure of ADFM. i) Hierarchical aggregation unit aggregates raw sequences to remove duplicate behaviors. ii) Behavior selection unit refines each behavior and separates useful and useless behaviors. iii) Interest extraction unit and adversarial interest extraction unit capture user interests from useful and useless behaviors respectively.}
  \label{fig:network}
\end{figure}

The aim of ADFM is to identify duplicate and useless behaviors from the long-term user behavior sequence and retain the most useful k behaviors, where k is a hyper-parameter. Based on the filtered sequences, user's long-term interests are extracted and fed into the CTR model. The model structure of ADFM is shown in Figure~\ref{fig:network} and the main modules are as follows:

{\bfseries Hierarchical Aggregation Unit (HAU).} HAU uses a hierarchical aggregation representation to compress duplicate behaviors in raw sequences. For each sequence, it groups behaviors by their value (e.g., item id) and only keeps the unique behaviors. The representation is essentially a set instead of a sequence. Keeping unique behaviors compresses raw sequences significantly whereas it may cause the loss of sequential information and can not distinguish between the behaviors occurred once and multiple times. To preserve the original information as much as possible, we add two aggregation statistics (i.e., occurrence number and the maximum timestamp of behavior) and transform them into sparse features by binning. Each behavior is represented as a tuple $<behavior\_id, number, timestamp>$. Take a sequence with behavior type $click$ and behavior target $shop$ for example, the aggregated sequence can be: 
\begin{multline}
    s_{click,shop}^{u} = [<shop\_id_{1}, click\_num_{1}, timestamp_{1}> \\
    , ..., <shop\_id_{m}, click\_num_{m}, timestamp_{m}>]
\end{multline}

{\bfseries Behavior Selection Unit (BSU).} The sequences aggregated by HAU will be fed into embedding layer and output the embedding vectors of the sequences $ A_{t,v}^{u}=(e_{t,v}^{1},...,e_{t,v}^{i},...,e_{t,v}^{m}) $, where $m$ is the length of aggregated sequences. Here we stack $e_{t,v}^{i} \in R^{D}$ together into a matrix $E_{t,v}^{u} \in R^{m \times D}$. There have been no duplicate behaviors in aggregated sequences but still have a lot of useless behaviors. BSU aims to select top-k useful behaviors from aggregated sequences. 

Firstly, to capture the interactions between behaviors in aggregated sequence, we use the self-attention mechanism~\cite{vaswani2017attention,bert_arxiv} to obtain a new embedding vector for each behavior:
\begin{align}
\begin{split}
    E_{t, v}^{u, att} {} &= Attention(E_{t, v}^{u}W^{Q},E_{t, v}^{u}W^{K},E_{t, v}^{u}W^{V}) \\
    &= Softmax(\frac{E_{t, v}^{u}W^{Q}(E_{t, v}^{u}W^{K})^{T}}{\sqrt{|D|}})E_{t, v}^{u}W^{V}
\end{split}
\end{align}
where $W^{Q}$, $W^{K}$ and $W^{V}$ are linear matrices. 

$E_{t, v}^{u}$ is then fed into a scoring gate and output selection scores $G_{t,v}$, which reflects the importance of sequence behaviors. \begin{equation}
    G_{t,v} = F(E_{t, v}^{u, att})
\end{equation}
where $F(.)$ is implemented as a feed forward neural network which is jointly trained with CTR model. We multiply $E_{t, v}^{u, att}$ by $G_{t,v}$ to adjust embedding vector of each behavior and obtain the embedding vectors of selected behaviors by:
\begin{equation}
    E_{t,v}^{u,s} = Filter(G_{t,v},\ G_{t,v}E_{t, v}^{u, att},\ k),
\end{equation}
where the function $Filter(score, embedding, k)$ sorts sequence behaviors by score and selects top-k behaviors.

{\bfseries Interest Extraction Unit (IEU).} IEU adopts multi-head attention~\cite{vaswani2017attention,bert_arxiv} to extract user interest from the top-k behaviors:
\begin{equation}
    head_{i} = Softmax(\frac{E_{target}W^{C}(E_{t,v}^{u,s}W^{S})^{T}}{\sqrt{|D|}})E_{t,v}^{u,s}W^{E}
\end{equation}
where $W^{C}$, $W^{S}$, $W^{E}$ are linear matrices, $E_{target}$ represents the embedding of target item and $head_{i}$ is the $i$th head in multi-head attention. \sloppy The long-term user interest is represented as $concat(head_{1};...;head_{n})$ and then fed into MLP layer for CTR prediction. The CTR loss based on the top-k behaviors is denoted as $Loss_{s}$ and calculated by Eq. (\ref{eqn:ctr_loss}).

Ideally, CTR model can propagate gradient back to BSU and adjust selection score, which can guide the training of BSU. However, the optimization goal of model loss is the accuracy of CTR prediction, not the usefulness of the behaviors selected by BSU. It is hard to guarantee that the selected behaviors are top k most useful. To supervise the learning of BSU more directly, we propose an adversarial filtering mechanism, which consists of an Adversarial Interest Extraction Unit (AIEU) and an adversarial training schema.

{\bfseries Adversarial Interest Extraction Unit}. The input of this module is the remaining m-k behaviors, and the model structure is the same as that of IEU. Since the distribution of useful behaviors and useless behaviors are quite different, the parameters of AIEU are not shared with IEU. AIEU aims to identify potential user interests from the remaining m-k actions and feed the interests into MLP layer for CTR prediction. The CTR loss based on the remaining m-k behaviors is denoted as $Loss_{\overline{s}}$. Notice that AIEU is only used in offline training while IEU is used in both training and inference.

If the top-k behaviors selected by BSU are always useful, $Loss_{s}$ should be small and $Loss_{\overline{s}}$ should be large. We optimize the following minimization problem:
\begin{equation}\label{eq:loss_joint_bs}
    \min \  loss_s - loss_{\overline{s}}.
\end{equation}
The loss will force BSU to select the top-k useful behaviors while the relatively useless m-k behaviors flow into AIEU.

There is a problem in directly optimizing this loss: When the parameters of BSU update during model training, the distribution of the top-k and the remaining m-k behaviors may change drastically, which makes it difficult for IEU and AIEU to converge. The convergence of the CTR loss becomes difficult meanwhile. Inspired by the success of adversarial learning~\cite{gan,chang2019game,MEED}, we propose an adversarial training schema to accelerate the optimization of ADFM.

We first fix BSU and learn IEU and AIEU with the goal of extracting user interests from both the top-k behaviors and the remaining m-k behaviors. The optimization problem is defined as follows:
\begin{equation}
    \min_{IEU,AIEU} \  Loss_{s}+Loss_{\overline{s}},
\end{equation}
Then we fix IEU and AIEU and optimize BSU with the following minimization problem:
\begin{equation}
     \min_{BSU} \  Loss_{s}-Loss_{\overline{s}}.
\end{equation}
The training process loops until it satisfies the stopping criterion (e.g., the changes of $Loss_{s}$ and $Loss_{\overline{s}}$ are small).

\section{Experiments}
In this Section we present the experimental setup and conduct experiments to evaluate the performance and cost of our model.

\subsection{Experimental setup}

{\bfseries Datasets.}
Two public datasets and an industrial dataset are used:

{\bfseries Amazon dataset\footnote{\url{https://nijianmo.github.io/amazon/index.html}}}. It is a commonly used benchmark, which is composed of product reviews and meta-data from Amazon~\cite{amazon}. We use the Books category of Amazon dataset, including 51 million records, 1.5 million users and 2.9 million items from 1252 categories. We treat review as click behavior. For long-term use behavior model, we filter the samples whose sequence length is shorter than 20.

{\bfseries Taobao dataset\footnote{\url{https://tianchi.aliyun.com/dataset/dataDetail?dataId=649}}}. It is a collection of user behaviors from Taobao’s recommender system~\cite{zhu2018learning} and contains 89 million records, 1 million users and 4 million items from 9407 categories. We only use click behaviors of each user.

{\bfseries Industrial dataset}. It is collected from our online advertising system in April 2021. The data collector has signed a contract with each user and obtained their consent to use the data. The dataset contains 4.3 billion records, 11.5 million users, 16.4 million items from 5000 categories. Each record contains historical behavior sequences with several behavior types (e.g., impression, click, add to chart and pay) from the preceding 180 days.

{\bfseries Competitors.} We compare ADFM with these competitors:
\begin{itemize}[leftmargin=*]
\item {{\bfseries DNN}} is a base DNN model presented in Section 2.1.

\item {{\bfseries DIN}~\cite{DIN}} proposes an attention mechanism to represent the user interests w.r.t. candidates. 

\item {{\bfseries DIEN}}~\cite{DIEN} uses GRU to model user interest evolution.

\item {{\bfseries ComicRec}}~\cite{2020ComicRec} proposes a comprehensive framework which integrates the controllability and multi-interest components. 

\item {{\bfseries MIMN}}~\cite{MIMN} proposes a memory network-based  model to capture multiple channels of user interest drifting for long-term user behavior modeling.

\item {{\bfseries SIM}}~\cite{SIM} is a search-based user interest model for long-term user behavior modeling. We will only compare the hard-search strategy as it is adopted in their online system.
\end{itemize}

Among these models, DNN, DIN, DIEN, ComicRec are designed for short-term user behavior, and MIMN, SIM are for long-term.

{\bfseries Parameter Configuration.}
For short-term user behavior models, the maximum sequence length is set to 20 and 100 in public datasets and industrial dataset respectively. For long-term ones, all historical behaviors are used. For ADFM, the number of selected behaviors is 20 and 100 in public datasets and industrial dataset respectively. MLP shape is 256 * 128 * 64 and embedding dimension $D$ is 16. . The head number in multi-head attention is set to 2. Optimization algorithm is Adam~\cite{adam} with learning rate 0.001. 

{\bfseries Evaluation Metrics.} We use $AUC$, $ GAUC_{user} $ and $ GAUC_{req} $ to evaluate these models. $AUC$ is a widely used metric to measure model effectiveness. $ GAUC_{user} $ and $ GAUC_{req} $ is group AUC~\cite{ocpc} calculated at the level of user and request respectively. We only use GAUC in industrial dataset since the number of groups is relatively small in public datasets.

\begin{table}
  \caption{Results on public datasets and industrial dataset}
  \label{tab:public and industrial result}
  \begin{tabular}{l|ll|lll}
    \toprule
    \multirow{2}*{Models} & Amazon & Taobao &
    \multicolumn{3}{c}{Industrial dataset}\\ 
    \cline{2-6}
    & $AUC$ &$AUC$& $AUC$& $GAUC_{user} $& $GAUC_{req} $\\
    \midrule
    DNN & 0.8355& 0.8714& 0.7001& 0.6354& 0.6083\\
    DIN & 0.8386& 0.8804& 0.7019& 0.6367& 0.6094\\
    DIEN & 0.8442& 0.9032& 0.7034& 0.6379& 0.6106\\
    ComicRec & 0.8458& 0.9012& 0.7037& 0.6381& 0.6109\\
    MIMN & 0.8494& 0.9144& 0.7046& 0.6392& 0.6131\\
    SIM & 0.8461& 0.9360& 0.7047& 0.6390& 0.6128\\
    ADFM & \textbf{0.8574}& \textbf{0.9462}& \textbf{0.7094}& \textbf{0.6428}& \textbf{0.6161}\\ \hline
\end{tabular}
\end{table}

\subsection{Performance evaluation}
As shown in Table~\ref{tab:public and industrial result}, MIMN, SIM and ADFM outperform DNN, DIN, DIEN and ComicRec since long-term user behavior sequence brings much new information. ADFM outperforms the other long-term user behavior models with a significant AUC gain of at least 0.1 in Amazon and Taobao dataset. ADFM also achieves an $AUC$ gain of 0.0047, a $ GAUC_{user} $ gain of 0.0038 and a $ GAUC_{req} $ gain of 0.0033 over the best competitor SIM in industrial dataset. In online advertisement systems with huge traffic, even 0.001 absolute AUC gain is a significant improvement~\cite{DIN}.

We conduct online A/B testing experiments to evaluate our model in our advertising system. The experiment lasts for 12 days and ADFM achieves 4.7$\%$ CTR and 3.1$\%$ RPM gain compared to SIM. To deploy ADFM, we keep the top-k behaviors selected by BSU and upload them to online cache. Online system fetches the selected sequences per user and feeds them to ADFM for CTR prediction. For SIM, we store user-category level behavior sequences in online cache and look up one sequence for each candidate item.

\subsection{Storage and latency cost}
During online inference, ADFM only stores the top-k behaviors online and storage cost is proportional to k. We analyze the contribution of HAU and BSU on cutting storage cost in industrial dataset. HAU compresses sequence length by $60\%$ and BSU reduces length by $27\%$ further. For online latency, MIMN maintains a fixed-length memory and the latency is nearly the same as that of ADFM if memory size equals to k. As sequence length increases, the latency of ADFM remains constant whereas SIM continues to increase.

\begin{table}
  \caption{Ablation study}
  \label{tab:ablation study}
  \begin{tabular}{l|ll lll}
    \toprule
    Model & AUC \\
    \midrule
    $DNN$ & 0.7001\\
    $DNN_{long}$ & 0.7028\\
    $DNN_{long}+HAU$ & 0.7035\\
    $DNN_{long}+HAU+BSU$ & 0.7061\\
    $DNN_{long}+HAU+BSU+adversarial\ learning (ADFM)$ & \textbf{0.7094}\\
    $DNN_{long}+HAU+ComicRec$ & 0.7074\\
  \bottomrule
\end{tabular}
\end{table}

\subsection{Ablation study}

We conduct an ablation study on industrial dataset to evaluate the effect of key modules of ADFM. As shown in Table~\ref{tab:ablation study}), $DNN_{long}$ is feeding long-term sequences into base CTR model and processing by sum-pooling. It is no surprise that it outperforms $DNN$ by introducing new information. $DNN_{long}+HAU$ performs slightly better than $DNN_{long}$ and demonstrates the effectiveness of hierarchical aggregation representation. Introducing BSU for selecting useful behaviors brings great improvements and adversarial learning strengthens the power of BSU by providing more supervision information for the learning of BSU. In addition, we evaluate the performance of $ComicRec$ on top of $DNN_{long}+HAU$. The AUC is higher than that of $DNN_{long}+HAU+BSU$ but less than that of ADFM, which indicates that BSU may be hard to optimize and adversarial learning alleviates the problem.

\section{Conclusion}
In this paper, we propose a novel adversarial filtering model on long-term user behavior sequences. We use a hierarchical aggregation representation to remove duplicate behaviors. To filter useless behaviors, our model scores and selects top-k useful behaviors with the help of an adversarial filtering mechanism. The results of offline and online A/B experiments demonstrate that our model achieves significant improvements over the competitors.

\bibliographystyle{ACM-Reference-Format}
\bibliography{sample-base}


\begin{thebibliography}{18}


\ifx \showCODEN    \undefined \def \showCODEN     #1{\unskip}     \fi
\ifx \showDOI      \undefined \def \showDOI       #1{#1}\fi
\ifx \showISBNx    \undefined \def \showISBNx     #1{\unskip}     \fi
\ifx \showISBNxiii \undefined \def \showISBNxiii  #1{\unskip}     \fi
\ifx \showISSN     \undefined \def \showISSN      #1{\unskip}     \fi
\ifx \showLCCN     \undefined \def \showLCCN      #1{\unskip}     \fi
\ifx \shownote     \undefined \def \shownote      #1{#1}          \fi
\ifx \showarticletitle \undefined \def \showarticletitle #1{#1}   \fi
\ifx \showURL      \undefined \def \showURL       {\relax}        \fi
\providecommand\bibfield[2]{#2}
\providecommand\bibinfo[2]{#2}
\providecommand\natexlab[1]{#1}
\providecommand\showeprint[2][]{arXiv:#2}

\bibitem[\protect\citeauthoryear{Cen, Zhang, Zou, Zhou, Yang, and Tang}{Cen
  et~al\mbox{.}}{2020}]%
        {2020ComicRec}
\bibfield{author}{\bibinfo{person}{Yukuo Cen}, \bibinfo{person}{Jianwei Zhang},
  \bibinfo{person}{Xu Zou}, \bibinfo{person}{Chang Zhou},
  \bibinfo{person}{Hongxia Yang}, {and} \bibinfo{person}{Jie Tang}.}
  \bibinfo{year}{2020}\natexlab{}.
\newblock \showarticletitle{Controllable multi-interest framework for
  recommendation}. In \bibinfo{booktitle}{\emph{Proceedings of the 26th ACM
  SIGKDD International Conference on Knowledge Discovery \& Data Mining}}.
  \bibinfo{publisher}{Association for Computing Machinery},
  \bibinfo{address}{New York, NY, USA}, \bibinfo{pages}{2942--2951}.
\newblock


\bibitem[\protect\citeauthoryear{Chang, Zhang, Yu, and Jaakkola}{Chang
  et~al\mbox{.}}{2019}]%
        {chang2019game}
\bibfield{author}{\bibinfo{person}{Shiyu Chang}, \bibinfo{person}{Yang Zhang},
  \bibinfo{person}{Mo Yu}, {and} \bibinfo{person}{Tommi Jaakkola}.}
  \bibinfo{year}{2019}\natexlab{}.
\newblock \showarticletitle{A game theoretic approach to class-wise selective
  rationalization}. In \bibinfo{booktitle}{\emph{Advances in Neural Information
  Processing Systems}}. \bibinfo{pages}{10055--10065}.
\newblock


\bibitem[\protect\citeauthoryear{Devlin, Chang, Lee, and Toutanova}{Devlin
  et~al\mbox{.}}{2018}]%
        {bert_arxiv}
\bibfield{author}{\bibinfo{person}{Jacob Devlin}, \bibinfo{person}{Ming-Wei
  Chang}, \bibinfo{person}{Kenton Lee}, {and} \bibinfo{person}{Kristina
  Toutanova}.} \bibinfo{year}{2018}\natexlab{}.
\newblock \bibinfo{title}{BERT: pre-training of deep bidirectional transformers
  for language understanding}.
\newblock
\newblock
\urldef\tempurl%
\url{https://doi.org/10.48550/ARXIV.1810.04805}
\showDOI{\tempurl}


\bibitem[\protect\citeauthoryear{Goodfellow, Pouget-Abadie, Mirza, Xu,
  Warde-Farley, Ozair, Courville, and Bengio}{Goodfellow et~al\mbox{.}}{2014}]%
        {gan}
\bibfield{author}{\bibinfo{person}{Ian Goodfellow}, \bibinfo{person}{Jean
  Pouget-Abadie}, \bibinfo{person}{Mehdi Mirza}, \bibinfo{person}{Bing Xu},
  \bibinfo{person}{David Warde-Farley}, \bibinfo{person}{Sherjil Ozair},
  \bibinfo{person}{Aaron Courville}, {and} \bibinfo{person}{Yoshua Bengio}.}
  \bibinfo{year}{2014}\natexlab{}.
\newblock \showarticletitle{Generative adversarial nets}. In
  \bibinfo{booktitle}{\emph{Advances in Neural Information Processing
  Systems}}, \bibfield{editor}{\bibinfo{person}{Z.~Ghahramani},
  \bibinfo{person}{M.~Welling}, \bibinfo{person}{C.~Cortes},
  \bibinfo{person}{N.~Lawrence}, {and} \bibinfo{person}{K.Q. Weinberger}}
  (Eds.), Vol.~\bibinfo{volume}{27}. \bibinfo{publisher}{Curran Associates,
  Inc.}
\newblock


\bibitem[\protect\citeauthoryear{Kingma and Ba}{Kingma and Ba}{2014}]%
        {adam}
\bibfield{author}{\bibinfo{person}{Diederik~P Kingma} {and}
  \bibinfo{person}{Jimmy Ba}.} \bibinfo{year}{2014}\natexlab{}.
\newblock \showarticletitle{Adam: a method for stochastic optimization}.
\newblock \bibinfo{journal}{\emph{arXiv preprint arXiv:1412.6980}}
  (\bibinfo{year}{2014}).
\newblock


\bibitem[\protect\citeauthoryear{Liang, Bai, Cao, Bai, and Wang}{Liang
  et~al\mbox{.}}{2020}]%
        {MEED}
\bibfield{author}{\bibinfo{person}{Jian Liang}, \bibinfo{person}{Bing Bai},
  \bibinfo{person}{Yuren Cao}, \bibinfo{person}{Kun Bai}, {and}
  \bibinfo{person}{Fei Wang}.} \bibinfo{year}{2020}\natexlab{}.
\newblock \showarticletitle{Adversarial infidelity learning for model
  interpretation}. In \bibinfo{booktitle}{\emph{Proceedings of the 26th ACM
  SIGKDD International Conference on Knowledge Discovery \& Data Mining}}.
  \bibinfo{pages}{286--296}.
\newblock


\bibitem[\protect\citeauthoryear{Ni, Li, and McAuley}{Ni et~al\mbox{.}}{2019}]%
        {amazon}
\bibfield{author}{\bibinfo{person}{Jianmo Ni}, \bibinfo{person}{Jiacheng Li},
  {and} \bibinfo{person}{Julian McAuley}.} \bibinfo{year}{2019}\natexlab{}.
\newblock \showarticletitle{Justifying recommendations using distantly-labeled
  reviews and fine-grained aspects}. In \bibinfo{booktitle}{\emph{Proceedings
  of the 2019 Conference on Empirical Methods in Natural Language Processing
  and the 9th International Joint Conference on Natural Language Processing
  (EMNLP-IJCNLP)}}. \bibinfo{pages}{188--197}.
\newblock


\bibitem[\protect\citeauthoryear{Pi, Bian, Zhou, Zhu, and Gai}{Pi
  et~al\mbox{.}}{2019}]%
        {MIMN}
\bibfield{author}{\bibinfo{person}{Qi Pi}, \bibinfo{person}{Weijie Bian},
  \bibinfo{person}{Guorui Zhou}, \bibinfo{person}{Xiaoqiang Zhu}, {and}
  \bibinfo{person}{Kun Gai}.} \bibinfo{year}{2019}\natexlab{}.
\newblock \showarticletitle{Practice on long sequential user behavior modeling
  for click-through rate prediction}. In \bibinfo{booktitle}{\emph{Proceedings
  of the 25th ACM SIGKDD International Conference on Knowledge Discovery \&
  Data Mining}}. \bibinfo{publisher}{Association for Computing Machinery},
  \bibinfo{address}{New York, NY, USA}, \bibinfo{pages}{2671--2679}.
\newblock


\bibitem[\protect\citeauthoryear{Pi, Zhou, Zhang, Wang, Ren, Fan, Zhu, and
  Gai}{Pi et~al\mbox{.}}{2020}]%
        {SIM}
\bibfield{author}{\bibinfo{person}{Qi Pi}, \bibinfo{person}{Guorui Zhou},
  \bibinfo{person}{Yujing Zhang}, \bibinfo{person}{Zhe Wang},
  \bibinfo{person}{Lejian Ren}, \bibinfo{person}{Ying Fan},
  \bibinfo{person}{Xiaoqiang Zhu}, {and} \bibinfo{person}{Kun Gai}.}
  \bibinfo{year}{2020}\natexlab{}.
\newblock \showarticletitle{Search-based user interest modeling with lifelong
  sequential behavior data for click-through rate prediction}. In
  \bibinfo{booktitle}{\emph{Proceedings of the 29th ACM International
  Conference on Information \& Knowledge Management}}.
  \bibinfo{publisher}{Association for Computing Machinery},
  \bibinfo{pages}{2685--2692}.
\newblock


\bibitem[\protect\citeauthoryear{Qin, Zhang, Wu, Jin, Fang, and Yu}{Qin
  et~al\mbox{.}}{2020}]%
        {UBR}
\bibfield{author}{\bibinfo{person}{Jiarui Qin}, \bibinfo{person}{Weinan Zhang},
  \bibinfo{person}{Xin Wu}, \bibinfo{person}{Jiarui Jin},
  \bibinfo{person}{Yuchen Fang}, {and} \bibinfo{person}{Yong Yu}.}
  \bibinfo{year}{2020}\natexlab{}.
\newblock \showarticletitle{User behavior retrieval for click-through rate
  prediction}. In \bibinfo{booktitle}{\emph{Proceedings of the 43rd
  International ACM SIGIR Conference on Research and Development in Information
  Retrieval}}. \bibinfo{publisher}{Association for Computing Machinery},
  \bibinfo{address}{New York, NY, USA}, \bibinfo{pages}{2347–2356}.
\newblock
\showISBNx{9781450380164}


\bibitem[\protect\citeauthoryear{Ren, Qin, Fang, Zhang, Zheng, Bian, Zhou, Xu,
  Yu, Zhu, et~al\mbox{.}}{Ren et~al\mbox{.}}{2019}]%
        {HPMN}
\bibfield{author}{\bibinfo{person}{Kan Ren}, \bibinfo{person}{Jiarui Qin},
  \bibinfo{person}{Yuchen Fang}, \bibinfo{person}{Weinan Zhang},
  \bibinfo{person}{Lei Zheng}, \bibinfo{person}{Weijie Bian},
  \bibinfo{person}{Guorui Zhou}, \bibinfo{person}{Jian Xu},
  \bibinfo{person}{Yong Yu}, \bibinfo{person}{Xiaoqiang Zhu}, {et~al\mbox{.}}}
  \bibinfo{year}{2019}\natexlab{}.
\newblock \showarticletitle{Lifelong sequential modeling with personalized
  memorization for user response prediction}. In
  \bibinfo{booktitle}{\emph{Proceedings of the 42nd International ACM SIGIR
  Conference on Research and Development in Information Retrieval}}.
  \bibinfo{pages}{565--574}.
\newblock


\bibitem[\protect\citeauthoryear{Sun, Liu, Wu, Pei, Lin, Ou, and Jiang}{Sun
  et~al\mbox{.}}{2019}]%
        {bert4rec}
\bibfield{author}{\bibinfo{person}{Fei Sun}, \bibinfo{person}{Jun Liu},
  \bibinfo{person}{Jian Wu}, \bibinfo{person}{Changhua Pei},
  \bibinfo{person}{Xiao Lin}, \bibinfo{person}{Wenwu Ou}, {and}
  \bibinfo{person}{Peng Jiang}.} \bibinfo{year}{2019}\natexlab{}.
\newblock \showarticletitle{BERT4Rec: sequential recommendation with
  bidirectional encoder representations from transformer}. In
  \bibinfo{booktitle}{\emph{Proceedings of the 28th ACM International
  Conference on Information and Knowledge Management}} (Beijing, China)
  \emph{(\bibinfo{series}{CIKM '19})}. \bibinfo{publisher}{Association for
  Computing Machinery}, \bibinfo{address}{New York, NY, USA},
  \bibinfo{pages}{1441–1450}.
\newblock
\showISBNx{9781450369763}


\bibitem[\protect\citeauthoryear{Vaswani, Shazeer, Parmar, Uszkoreit, Jones,
  Gomez, Kaiser, and Polosukhin}{Vaswani et~al\mbox{.}}{2017}]%
        {vaswani2017attention}
\bibfield{author}{\bibinfo{person}{Ashish Vaswani}, \bibinfo{person}{Noam
  Shazeer}, \bibinfo{person}{Niki Parmar}, \bibinfo{person}{Jakob Uszkoreit},
  \bibinfo{person}{Llion Jones}, \bibinfo{person}{Aidan~N Gomez},
  \bibinfo{person}{\L~ukasz Kaiser}, {and} \bibinfo{person}{Illia Polosukhin}.}
  \bibinfo{year}{2017}\natexlab{}.
\newblock \showarticletitle{Attention is all you need}. In
  \bibinfo{booktitle}{\emph{Advances in Neural Information Processing
  Systems}}, \bibfield{editor}{\bibinfo{person}{I.~Guyon},
  \bibinfo{person}{U.~Von Luxburg}, \bibinfo{person}{S.~Bengio},
  \bibinfo{person}{H.~Wallach}, \bibinfo{person}{R.~Fergus},
  \bibinfo{person}{S.~Vishwanathan}, {and} \bibinfo{person}{R.~Garnett}}
  (Eds.), Vol.~\bibinfo{volume}{30}. \bibinfo{publisher}{Curran Associates,
  Inc.}
\newblock


\bibitem[\protect\citeauthoryear{Zhang, Qin, Guo, Tang, and He}{Zhang
  et~al\mbox{.}}{2021}]%
        {DeepForCtr}
\bibfield{author}{\bibinfo{person}{Weinan Zhang}, \bibinfo{person}{Jiarui Qin},
  \bibinfo{person}{Wei Guo}, \bibinfo{person}{Ruiming Tang}, {and}
  \bibinfo{person}{Xiuqiang He}.} \bibinfo{year}{2021}\natexlab{}.
\newblock \showarticletitle{Deep learning for click-through rate estimation}.
  In \bibinfo{booktitle}{\emph{Proceedings of the Thirtieth International Joint
  Conference on Artificial Intelligence, {IJCAI-21}}},
  \bibfield{editor}{\bibinfo{person}{Zhi-Hua Zhou}} (Ed.).
  \bibinfo{publisher}{International Joint Conferences on Artificial
  Intelligence Organization}, \bibinfo{pages}{4695--4703}.
\newblock
\urldef\tempurl%
\url{https://doi.org/10.24963/ijcai.2021/636}
\showURL{%
\tempurl}
\newblock
\shownote{Survey Track.}


\bibitem[\protect\citeauthoryear{Zhou, Mou, Fan, Pi, Bian, Zhou, Zhu, and
  Gai}{Zhou et~al\mbox{.}}{2019}]%
        {DIEN}
\bibfield{author}{\bibinfo{person}{Guorui Zhou}, \bibinfo{person}{Na Mou},
  \bibinfo{person}{Ying Fan}, \bibinfo{person}{Qi Pi}, \bibinfo{person}{Weijie
  Bian}, \bibinfo{person}{Chang Zhou}, \bibinfo{person}{Xiaoqiang Zhu}, {and}
  \bibinfo{person}{Kun Gai}.} \bibinfo{year}{2019}\natexlab{}.
\newblock \showarticletitle{Deep interest evolution network for click-through
  rate prediction}. In \bibinfo{booktitle}{\emph{Proceedings of the AAAI
  conference on artificial intelligence}}, Vol.~\bibinfo{volume}{33}.
  \bibinfo{pages}{5941--5948}.
\newblock


\bibitem[\protect\citeauthoryear{Zhou, Zhu, Song, Fan, Zhu, Ma, Yan, Jin, Li,
  and Gai}{Zhou et~al\mbox{.}}{2018}]%
        {DIN}
\bibfield{author}{\bibinfo{person}{Guorui Zhou}, \bibinfo{person}{Xiaoqiang
  Zhu}, \bibinfo{person}{Chenru Song}, \bibinfo{person}{Ying Fan},
  \bibinfo{person}{Han Zhu}, \bibinfo{person}{Xiao Ma},
  \bibinfo{person}{Yanghui Yan}, \bibinfo{person}{Junqi Jin},
  \bibinfo{person}{Han Li}, {and} \bibinfo{person}{Kun Gai}.}
  \bibinfo{year}{2018}\natexlab{}.
\newblock \showarticletitle{Deep interest network for click-through rate
  prediction}. In \bibinfo{booktitle}{\emph{Proceedings of the 24th ACM SIGKDD
  International Conference on Knowledge Discovery \& Data Mining}}.
  \bibinfo{pages}{1059--1068}.
\newblock


\bibitem[\protect\citeauthoryear{Zhu, Jin, Tan, Pan, Zeng, Li, and Gai}{Zhu
  et~al\mbox{.}}{2017}]%
        {ocpc}
\bibfield{author}{\bibinfo{person}{Han Zhu}, \bibinfo{person}{Junqi Jin},
  \bibinfo{person}{Chang Tan}, \bibinfo{person}{Fei Pan},
  \bibinfo{person}{Yifan Zeng}, \bibinfo{person}{Han Li}, {and}
  \bibinfo{person}{Kun Gai}.} \bibinfo{year}{2017}\natexlab{}.
\newblock \showarticletitle{Optimized cost per click in Taobao display
  advertising}. In \bibinfo{booktitle}{\emph{Proceedings of the 23rd ACM SIGKDD
  International Conference on Knowledge Discovery and Data Mining}} (Halifax,
  NS, Canada) \emph{(\bibinfo{series}{KDD '17})}.
  \bibinfo{publisher}{Association for Computing Machinery},
  \bibinfo{address}{New York, NY, USA}, \bibinfo{pages}{2191–2200}.
\newblock
\showISBNx{9781450348874}


\bibitem[\protect\citeauthoryear{Zhu, Li, Zhang, Li, He, Li, and Gai}{Zhu
  et~al\mbox{.}}{2018}]%
        {zhu2018learning}
\bibfield{author}{\bibinfo{person}{Han Zhu}, \bibinfo{person}{Xiang Li},
  \bibinfo{person}{Pengye Zhang}, \bibinfo{person}{Guozheng Li},
  \bibinfo{person}{Jie He}, \bibinfo{person}{Han Li}, {and}
  \bibinfo{person}{Kun Gai}.} \bibinfo{year}{2018}\natexlab{}.
\newblock \showarticletitle{Learning tree-based deep model for recommender
  systems}. In \bibinfo{booktitle}{\emph{Proceedings of the 24th ACM SIGKDD
  International Conference on Knowledge Discovery \& Data Mining}}.
  \bibinfo{publisher}{Association for Computing Machinery},
  \bibinfo{pages}{1079--1088}.
\newblock


\end{thebibliography}

\end{document}